\documentclass[reviewcopy]{elsarticle}

\usepackage[reviewcopy]{adndt}
\usepackage{longtable}

\usepackage{amsmath}

\usepackage{amssymb}

\biboptions{square,sort&compress}
\bibpunct[]{[}{]}{,}{n}{}{;}
\begin{document}

\begin{frontmatter}

\journal{Atomic Data and Nuclear Data Tables}


\title{Radiative rates for E1, E2, M1, and M2 transitions in the Br-like ions Sr IV, Y V, Zr VI, Nb VII, and Mo VIII}

  \author[One]{Kanti M. Aggarwal\fnref{}\corref{cor1}}
  \ead{K.Aggarwal@qub.ac.uk}

  \author[One]{Francis P. Keenan}


  \cortext[cor1]{Corresponding author.}

  \address[One]{Astrophysics Research Centre, School of Mathematics and Physics, Queen's University Belfast,\\Belfast BT7 1NN,
Northern Ireland, UK}


\date{16.12.2002} 

\begin{abstract}  
Energies and lifetimes are reported for the lowest 375  levels of five Br-like ions, namely Sr~IV, Y~V, Zr~VI, Nb~VII, and Mo~VIII,  mostly belonging to the 4s$^2$4p$^5$, 4s$^2$4p$^4$4$\ell$, 4s4p$^6$,  4s$^2$4p$^4$5$\ell$,  4s$^2$4p$^3$4d$^2$,  4s4p$^5$4$\ell$, and  4s4p$^5$5$\ell$  configurations. Extensive configuration interaction has been included and the general-purpose relativistic atomic structure package ({\sc grasp}) has been adopted for the calculations. Additionally,  radiative rates are listed among these levels for all E1, E2, M1, and M2 transitions. From a  comparison with  the measurements, the majority of our energy levels are assessed to be accurate to better than  2\%, although discrepancies between theory and experiment for a few are up to 6\%. An  accuracy assessment of the calculated  radiative rates (and lifetimes)  is more difficult, because no prior results exist for these ions. 

\end{abstract}

\end{frontmatter}




\newpage

\tableofcontents
\listofDtables
\listofDfigures
\vskip5pc


\section{Introduction}
Atomic data for  energy levels and  radiative decay rates (A-values) are required for the modelling of of plasmas, such as astrophysical and fusion. Generally, comparatively lighter ions ($Z  \le$ 30) are of relevance  to both astrophysical and fusion plasmas, but the heavier ones are mostly useful for the latter, mainly for  assessing  and controlling  the radiation loss. With the ongoing  ITER project, the need for atomic data for a wide range of  ions has become greater.  Therefore, in a recent paper, we \cite{w40} have reported atomic data for an important Br-like  tungsten ion, W XL. In this paper we provide similar results for five additional Br-like ions with 38 $\le$ Z $\le$ 42, i.e. Sr ~IV, Y~V, Zr~VI, Nb~VII, and Mo~VIII.

Laboratory measurements  for energy levels are available for all five Br-like ions of interest, but are limited to only a few levels, except for Sr IV. These measurements have been  compiled by the NIST (National Institute of Standards and Technology) team \cite{nist}, and are  available at their website {\tt http://physics.nist.gov/PhysRefData/ASD/levels\_form.html}. However, similar data for A-values  \cite{bch}  are limited to the  magnetic dipole (M1) and electric quadrupole (E2) transitions between the ground state levels (4s$^2$4p$^5$ $^2$P$^o_{3/2}$ -- $^2$P$^o_{1/2}$) of several Br-like ions, including those considered here,  insufficient for modelling applications. Therefore, in a recent paper Singh  at al.  \cite{sam} have calculated  energies and lifetimes for the lowest 31 levels, belonging to  the 4s$^2$4p$^5$, 4s$^2$4p$^4$4d and 4s4p$^6$  configurations, of the five Br-like ions listed above. Additionally, they also listed A-values for electric  dipole (E1) transitions, but only from the ground state  4s$^2$4p$^5$ $^2$P$^o_{3/2,1/2}$ to higher-lying levels. For the calculations, they adopted the GRASP code  \cite{grasp0} but included only limited CI (configuration interaction). However, these 31 levels are not the lowest, because many of the 4s$^2$4p$^4$5s  configuration intermix with these -- see Tables 1--5.  More importantly, their reported results cannot be reproduced. Discrepancies with our calculated energy levels, obtained with the same code, configurations and approximations, are only up to 0.15 Ryd, but are much  larger (up to 5 orders of magnitude) for  A-values and lifetimes.  In our earlier paper \cite{ak} we demonstrated that these discrepancies are due to the results of \cite{sam} being unreliable. However, in \cite{ak}  the atomic data are restricted to the 31 levels of the 4s$^2$4p$^5$, 4s$^2$4p$^4$4d and 4s4p$^6$  configurations, although calculations were performed for up to 3990 levels. Therefore, in this paper we report complete results among the lowest 375 levels of these ions.

\section{Energy levels}

For our  calculations we have adopted the  fully relativistic multi-configuration Dirac-Fock (MCDF) code,  originally developed by Grant  et al.  \cite{grasp0}. The code is  based on the $jj$ coupling scheme and includes higher-order relativistic corrections arising from the Breit interaction and QED (quantum electrodynamics) effects, which are important for the heavy ions considered here.  However, this initial version of the MCDF code  has undergone multiple revisions, and is now better known as  GRASP   \cite{grasp}. The version adopted in our work has been revised by one of its authors (Dr. P. H. Norrington),  is referred to as GRASP0, and  is freely available at  {\tt http://web.am.qub.ac.uk/DARC/}. It provides comparable results with other revisions, such as GRASP2K \cite{grasp2k, grasp2kk}.

In our calculations, extensive configuration interaction  (CI)  has been included among 39 configurations, namely 4s$^2$4p$^5$, 4s$^2$4p$^4$4d/4f, 4s4p$^6$, 4p$^6$4d/4f, 4s4p$^5$4d/4f, 4s$^2$4p$^3$4f$^2$/4d$^2$/4d4f, 4s$^2$4p$^2$4d$^3$, 4s$^2$4p4d$^4$,  4s$^2$4p$^2$4d$^2$4f, 4s4p$^3$4d$^3$, 4p$^5$4d$^2$, 3d$^9$4s$^2$4p$^5$4d/4f, 3d$^9$4s$^2$4p$^6$, 4s4p$^5$5$\ell$,   4p$^6$5$\ell$,  4s$^2$4p$^4$5$\ell$, and 3d$^9$4s$^2$4p$^5$5$\ell$. These configurations generate   3990 levels in total and have been carefully chosen, mainly based on the interacting range of their energies. The specific number of levels each configuration generates  and their  energy ranges are listed in Table 1 of our earlier paper \cite{ak}.  As  noted earlier for some Ni ions \cite{ni} the interacting energy ranges of the configurations significantly influence  the accuracy of the calculations. We also point out  that these configurations are not the same as included for Br-like W~XL \cite{w40}, although many are common. Furthermore,  additional  configurations, such as 3p$^5$3d$^{10}$4s$^2$4p$^6$, 3p$^5$3d$^{10}$4s$^2$4p$^5$4d and 3p$^5$3d$^{10}$4s$^2$4p$^5$4f, have also been tested but excluded from the final calculations, because their impact on the lower energy levels is insignificant,  mainly because they generate levels lying at much higher energies.  Nevertheless, we discuss their impact later in the section.  Finally, as in most of our earlier work,  we have adopted the option of  `extended average level',  in which a weighted (proportional to 2$j$+1) trace of the Hamiltonian matrix is minimised. This  yields results comparable to other options, such as `average level' -- see for example,  Aggarwal  et al.  for several ions of Kr \cite{kr} and Xe \cite{xe}.  The  ions considered here  have nuclear charges  between those of Kr and Xe.

In Tables 1--5 we list our lowest 375 energy levels for five Br-like ions with 38 $\le$ Z $\le$ 42. Comparisons with the measurements  compiled by NIST have already been discussed in our earlier work \cite{ak}, but only for  31 levels of the 4s$^2$4p$^5$, 4s$^2$4p$^4$4d and 4s4p$^6$  configurations. However, NIST energies are also available for a few other higher lying levels, particularly those of 4p$^4$5s for all ions, and of many more  configurations of Sr IV. Therefore, in Table A we compare our calculated energies with the NIST listings for the  common levels of  Sr~IV. For most  levels the theoretical energies agree with the NIST values within 2\%, but discrepancies for a few are up to 6\%, particularly among the lowest 60 levels in Table A.  

We would like to stress here that the $LSJ$ labels assigned  to the levels in Tables 1--5 are not always unique,  because  levels from different configurations mix strongly and sometimes the eigenvector of the same level/configuration dominates for two (or even three) levels. This is a common problem for all calculations, particularly when the CI is important as in the present case. However, the ambiguity in the configuration/level designation is only for a few levels in each ion. In Table B we list the mixing coefficients for the lowest 50 levels of  Sr~IV -- see in particular, levels 14, 18, 39, and 50, which are highly mixed. Hence,  the $LSJ$ designations provided in Tables 1--5  may sometimes be interchangeable, although  the associated $J^{\pi}$ values are definitive for all levels and ions. 

We now investigate if a larger CI expansion is helpful in improving the accuracy of our calculated energy levels. Adopting  the  {\em Flexible Atomic Code} ({\sc fac}) of Gu \cite{fac},  available from the website {\tt http://sprg.ssl.berkeley.edu/$\sim$mfgu/fac/}, we have performed a series of calculations with up to 12~137 levels. The additional levels  belong to the  4p$^6$6$\ell$,  4s4p$^5$6$\ell$, 4s$^2$4p$^4$6$\ell$,   4p$^6$7$\ell$,  4s4p$^5$7$\ell$, 4s$^2$4p$^4$7$\ell$, 4s$^2$4p$^3$5$\ell^2$, 4s4p$^4$5$\ell^2$, 3p$^5$3d$^{10}$4s$^2$4p$^6$, 3p$^5$3d$^{10}$4s$^2$4p$^5$4d, and 3p$^5$3d$^{10}$4s$^2$4p$^5$4f configurations. {\sc fac} is also a fully relativistic code and  results obtained for energy levels and A-values are  (generally) comparable to {\sc grasp}, as already shown for several other ions, see for example Aggarwal  et al. for Kr \cite{kr} and Xe \cite{xe} ions. This  larger calculation can  be more easily performed with this code  within a reasonable time frame of a few days, because of its higher efficiency. However, energies obtained from this larger FAC calculation agree with GRASP for the levels of the 4s$^2$4p$^5$, 4s$^2$4p$^4$4d and 4s4p$^6$  configurations,  within 0.01 Ryd for all ions, as  shown in our earlier work \cite{ak}. In Table C we make a similar comparison for all levels of the  4s$^2$4p$^5$, 4s4p$^6$, 4s$^2$4p$^4$4d/4f, and 4s$^2$4p$^4$5$\ell$  configurations of Sr~IV. These levels are not listed in the ascending order of energy, but in the order of the configurations for easier understanding. Furthermore,  the energy differences between the two independent calculations are listed in both magnitude and percentage for a ready reference. Although energy differences for most levels are below 1\%, discrepancies are up to 3.5\% for a few, particularly for those of the  $n$=4 configurations for which energies from FAC are invariably higher. Correspondingly energy differences between the NIST compilation and FAC calculation has increased for most levels, and therefore a larger expansion is not helpful. A closer agreement between theoretical and experimental energies for these levels may perhaps be achieved by the inclusion of an enormously large CI (involving over a million levels or configuration state functions, CSF), as  shown by Froese Fischer \cite{cff} for W~XL, where she performed extremely  large calculations for a limited set of levels. Unfortunately, for the greater  number of levels involved in the present work and for five ions, such calculations are not possible with our computational resources.

\section{Radiative rates}\label{sec.eqs}

The absorption oscillator strength ($f_{ij}$)  for a transition $i \to j$  is a dimensionless quantity  and is related to the radiative rate A$_{ji}$ (in s$^{-1}$) by the following expression:

\begin{equation}
f_{ij} = \frac{mc}{8{\pi}^2{e^2}}{\lambda^2_{ji}} \frac{{\omega}_j}{{\omega}_i}A_{ji}
 = 1.49 \times 10^{-16} \lambda^2_{ji} \frac{{\omega}_j}{{\omega}_i} A_{ji}
\end{equation}
where $m$ and $e$ are the electron mass and charge, respectively, $c$  the velocity of light,  $\lambda_{ji}$  the transition wavelength in $\rm \AA$, and $\omega_i$ and $\omega_j$  the statistical weights of the lower $i$ and upper $j$ levels, respectively.
Similarly, the oscillator strength $f_{ij}$ (dimensionless), A-values  and the line strength $S$ (in atomic units, 1 a.u. = 6.460$\times$10$^{-36}$ cm$^2$ esu$^2$) are related by the  standard equations given in \cite{w40}. 

In our calculations with the {\sc grasp} code the S- (and subsequently A- and f-)  values have been determined in both the length and velocity forms, equivalent to the Babushkin and Coulomb gauges in the relativistic nomenclature. In Tables 6--10 we present results  for transitions from the lowest three to higher excited levels, with the full tables  available online in the electronic version.   Included in these tables are the  transition (energies) wavelengths ($\lambda_{ij}$ in ${\rm \AA}$), radiative rates (A$_{ji}$ in s$^{-1}$), oscillator strengths ($f_{ij}$, dimensionless), and line strengths ($S$ in a.u.) for all 16~568 E1 transitions among the lowest 375 levels.  The listed wavelengths (and other parameters) are based on the Breit and QED-corrected theoretical energies,  given in Tables 1--5, where the {\em indices} used to represent the lower and upper levels of a transition are also defined. However,  for the  29~166 E2, 23~514  M1, and 20~476 M2 transitions,  only the A-values are included in Tables 6--10, because the corresponding results for f- or S-values can be  easily obtained using Eqs. (1-5) given in \cite{w40}. Since the length form is  considered to be comparatively  more accurate, results listed in Tables 6--10 are in this form alone. However, for accuracy assessment we have also listed the ratio R of the velocity and length forms for all E1 transitions.

For comparison purpose there are no available data in the literature for A-values. However, based on calculations with differing amount of CI,  codes, and the ratio of the forms in length and velocity, the accuracy of our A-values has  been assessed in our earlier paper \cite{ak}. It is estimated to be $\sim$20\% for transitions with f $\ge$ 0.01. 
However, that comparison was limited to transitions among the  31 levels of the 4s$^2$4p$^5$, 4s$^2$4p$^4$4d and 4s4p$^6$  configurations, and hence is not applicable to a wider range of transitions listed in Tables 6--10. For example, there are 2431 transitions of Sr~IV in Table 6 for which f(E1) $\ge$ 0.01, and $\sim$59\% differ by over 20\%. Their number reduces to 20\% if R is taken to be within 50\% and only 3\% transitions differ by over an order of magnitude -- see for example:  4--67, 5--66, and 6--65, all of which  have f $<$ 0.1. For weaker transitions the ratio R may be up to several orders of magnitude.

In Table D we compare our GRASP results for A- and f- values with the corresponding FAC calculations, for transitions from the lowest two to higher excited levels,  up to 49, of Sr~IV. Unfortunately, differences for a few transitions are up to a factor of 5. Most such transitions are comparatively weak (such as 1--11, 1--33 and 2--18), and larger differences for these are not surprising, because of differences in the methodology and the CI. However, the 1--32 transition is strong (f = 0.84) but the corresponding result from FAC is lower by a factor of three.  Inclusion of larger CI generally increases the accuracy of A-values. However, in the present case it may not be true as was discussed earlier in section 2 for energy levels. Finally,   there are no discrepancies with the Bi{\' e}mont et al.  \cite{bch} A-values for the M1 and E2 (4s$^2$4p$^5$) $^2$P$^o_{3/2}$ -- $^2$P$^o_{1/2}$ transitions.

\section{Lifetimes}

The lifetime $\tau$ of a level $j$ is determined  as follows:

\begin{equation}
{\tau}_j = \frac{1}{{\sum_{i}^{}} A_{ji}}.
\end{equation}
In Tables 1--5  we list lifetimes for the lowest 375 levels of Sr~IV, Y~V, Zr~VI, Nb~VII, and Mo~VIII.  The calculations of $\tau$  include A-values from all types of transitions, i.e. E1, E2, M1, and M2. Unfortunately, there are no measurements or other theoretical results available for comparison purposes. 

\section{Conclusions}

Adopting the {\sc grasp} code, energy levels and radiative rates (for E1, E2, M1, and M2 transitions)   for the lowest 375 levels of  5 Br-like ions (Sr~IV, Y~V, Zr~VI, Nb~VII, and Mo~VIII) have been presented. Additionally, lifetimes for these levels are also listed although no other data are available in the literature with which to compare our results. For most energy levels and for all ions, the accuracy of our results is estimated to be better than 2\%. This assessment is based on comparisons with the results  listed by  NIST. However, there is scope for improvement because for a few levels the discrepancies between theory and measurement are up to 6\%. Although length and velocity forms of the A-values have been discussed and comparisons with the FAC code with larger CI have been made, it is not possible to assign any accuracy estimate for the radiative data. Future calculations and measurements (particularly for lifetimes) will be helpful in the accuracy assessment of our listed data. 



\ack
KMA  is thankful to  AWE Aldermaston for financial support.  

\begin{appendix}

\def\thesection{} 

\section{Appendix A. Supplementary data}

Owing to space limitations, only parts of Tables 6--10  are presented here, the full tables being made available as supplemental material in conjunction with the electronic
publication of this work. Supplementary data associated with this article can be found, in the online version, at doi:nn.nnnn/j.adt.2015.nn.nnn.

\end{appendix}


\clearpage
\newpage

\renewcommand{\baselinestretch}{1.0}
\footnotesize
\begin{longtable}{@{\extracolsep\fill}rllllrllll@{}}
\caption{Comparison of energies (Ryd) for the lowest 221 levels of Sr IV. See Table 1 for the full listing of the levels.}
Index  & Configuration                   & Level                &  NIST      &  Present  & Index  & Configuration                  & Level                &  NIST      &  Present  \\
 \hline\\
\endfirsthead\\
\caption[]{(continued)}
Index  & Configuration                   & Level                &  NIST      &  Present  & Index  & Configuration                  & Level                &  NIST      &  Present  \\
\hline\\
\endhead
    1 &   4s$^2$4p$^5$  		 &    $^2$P$^o_{3/2 }$  &   0.00000  &  0.00000  &   71 &   4s$^2$4p$^4$5d		   &	$^4$D$  _{3/2 }$  &   2.98773  &  3.04428  \\	  
    2 &   4s$^2$4p$^5$  		 &    $^2$P$^o_{1/2 }$  &   0.08865  &  0.08515  &   72 &   4s$^2$4p$^4$($^3$P)4f	   &	$^4$G$^o_{11/2}$  &   3.05620  &  3.04851  \\	  
    3 &   4s4p$^6$			 &    $^2$S$  _{1/2 }$  &   1.37149  &  1.44244  &   73 &   4s$^2$4p$^4$($^3$P)4f	   &	$^2$F$^o_{7/2 }$  &   3.05986  &  3.05012  \\
    4 &   4s$^2$4p$^4$($^3$P)4d 	 &    $^4$D$  _{7/2 }$  &   1.70553  &  1.75452  &   74 &   4s$^2$4p$^4$($^3$P)4f	   &	$^2$F$^o_{5/2 }$  &   3.05919  &  3.05057  \\
    5 &   4s$^2$4p$^4$($^3$P)4d 	 &    $^4$D$  _{5/2 }$  &   1.70590  &  1.75612  &   75 &   4s$^2$4p$^4$($^3$P)4f	   &	$^2$G$^o_{9/2 }$  &   3.05985  &  3.05418  \\
    6 &   4s$^2$4p$^4$($^3$P)4d 	 &    $^4$D$  _{3/2 }$  &   1.71340  &  1.76432  &   76 &   4s$^2$4p$^4$5d		   &	$^4$D$  _{1/2 }$  &   3.02153  &  3.05480  \\
    7 &   4s$^2$4p$^4$($^3$P)4d 	 &    $^4$D$  _{1/2 }$  &   1.72338  &  1.77444  &   77 &   4s$^2$4p$^4$($^3$P)4f	   &	$^4$D$^o_{3/2 }$  &   3.06845  &  3.05433  \\
    8 &   4s$^2$4p$^4$($^3$P)4d 	 &    $^4$F$  _{9/2 }$  &   1.79574  &  1.85784  &   79 &   4s$^2$4p$^4$5d		   &	$^2$F$  _{7/2 }$  &   3.00694  &  3.07641  \\
\hline            								                	 
\end{longtable}   								   					       
			      							   					       

														       
\begin{flushleft}													       
{\small
NIST: http://physics.nist.gov/PhysRefData/ASD/levels\_form.html \\
GRASP: present calculations from the {\sc grasp} code with 3990 levels\\ 		       
}															       
\end{flushleft} 


\clearpage
\newpage

\begin{longtable}{@{\extracolsep\fill}rlll@{}}
\caption{Mixing coefficients (MC) for the lowest 50 levels of  Sr~IV.  Numbers outside and inside a bracket correspond to MC and the level, respectively. See Table 1 for definition of all levels.}
\\
Index  & Configuration           & Level              &  Mixing coefficients     \\
 \hline\\
\endfirsthead\\
\caption[]{(continued)}
Index  & Configuration           & Level              &  Mixing coefficients   \\
\hline\\
\endhead
    1   &   4s$^2$4p$^5$		     &  $^2$P$^o_{3/2 }$ &    0.97(  1) 				     \\
    2   &   4s$^2$4p$^5$		     &  $^2$P$^o_{1/2 }$ &    0.97(  2) 				     \\
    3   &   4s4p$^6$			     &  $^2$S$  _{1/2 }$ &    0.79(  3)-0.59( 39)			     \\
    4   &   4s$^2$4p$^4$($^3$P)4d	     &  $^4$D$  _{7/2 }$ &    0.97(  4) 				     \\
    5   &   4s$^2$4p$^4$($^3$P)4d	     &  $^4$D$  _{5/2 }$ &    0.96(  5) 				     \\
    6   &   4s$^2$4p$^4$($^3$P)4d	     &  $^4$D$  _{3/2 }$ &   -0.96(  6) 				     \\
    7   &   4s$^2$4p$^4$($^3$P)4d	     &  $^4$D$  _{1/2 }$ &   -0.97(  7) 				     \\
    8   &   4s$^2$4p$^4$($^3$P)4d	     &  $^4$F$  _{9/2 }$ &    0.96(  8)+0.23( 20)			     \\
\hline  											       								   					      
\end{longtable}

                               
\begin{flushleft}                                                                               
                               
{\small
                                                                               
}                                                                                               
                               
\end{flushleft} 

\clearpage
\newpage

\renewcommand{\baselinestretch}{1.0}
\footnotesize
\begin{longtable}{@{\extracolsep\fill}rllllrr@{}}
\caption{ Comparison of energies (Ryd) for some levels of Sr IV between the GRASP (3990 levels) and FAC (12~137 levels) calculations. See Table 1 for indices of all levels.}
Index  & Configuration                   & Level                &  GRASP  & FAC  & $\Delta$E  & Percentage Difference   \\ 
 \hline\\
\endfirsthead\\
\caption[]{(continued)}
Index  & Configuration                   & Level                &  GRASP  & FAC  & $\Delta$E  & Percentage Difference   \\ 
\hline\\
\endhead
    1 &   4s$^2$4p$^5$  		   &  $^2$P$^o_{3/2 }$   &   0.00000  &   0.00000  &	 0.000   &	0.0  \\
    2 &   4s$^2$4p$^5$  		   &  $^2$P$^o_{1/2 }$   &   0.08515  &   0.08812  &	-0.003   &     -3.5  \\
    3 &   4s4p$^6$			   &  $^2$S$  _{1/2 }$   &   1.44244  &   1.46391  &	-0.021   &     -1.5  \\
    4 &   4s$^2$4p$^4$($^3$P)4d 	   &  $^4$D$  _{7/2 }$   &   1.75452  &   1.78496  &	-0.030   &     -1.7  \\
    5 &   4s$^2$4p$^4$($^3$P)4d 	   &  $^4$D$  _{5/2 }$   &   1.75612  &   1.78502  &	-0.029   &     -1.6  \\
    6 &   4s$^2$4p$^4$($^3$P)4d 	   &  $^4$D$  _{3/2 }$   &   1.76432  &   1.79324  &	-0.029   &     -1.6  \\
    7 &   4s$^2$4p$^4$($^3$P)4d 	   &  $^4$D$  _{1/2 }$   &   1.77444  &   1.80563  &	-0.031   &     -1.8  \\
    8 &   4s$^2$4p$^4$($^3$P)4d 	   &  $^4$F$  _{9/2 }$   &   1.85784  &   1.88277  &	-0.025   &     -1.3  \\
\hline            								                	 
\end{longtable}   								   					       
			      							   					       

														       
\begin{flushleft}													       
{\small
}															       
\end{flushleft} 




\begin{longtable}{@{\extracolsep\fill}rrlllll@{}}
\caption{Comparison of radiative rates (A, s$^{-1}$) and oscillator strengths (f, dimensionless) for some transitions of Sr IV between the GRASP (3990 levels) and FAC (12~137 levels) calculations. The last column gives the raio of f(GRASP) and f(FAC).  $a{\pm}b \equiv a{\times}$10$^{{\pm}b}$. See Table 1 for definition of levels.}
\\
I  & J  & A(GRASP) & f (GRASP) & A(FAC)   & f (FAC) & R \\
 \hline\\
\endfirsthead\\
\caption[]{(continued)}
I  & J  & A(GRASP) & f (GRASP) & A(FAC)   & f (FAC) & R \\
\hline\\
\endhead
    1  &    3   &  3.546$+$08  &   1.061$-$02  &   4.074$+$08  &   1.183$-$02  &   0.90  \\
    1  &    5   &  1.024$+$07  &   6.197$-$04  &   1.181$+$07  &   6.920$-$04  &   0.90  \\
    1  &    6   &  2.556$+$06  &   1.022$-$04  &   4.138$+$06  &   1.602$-$04  &   0.64  \\
    1  &    7   &  4.257$+$05  &   8.415$-$06  &   1.383$+$06  &   2.640$-$05  &   0.32  \\
    1  &   10   &  8.106$+$07  &   4.138$-$03  &   1.047$+$08  &   5.188$-$03  &   0.80  \\
    1  &   11   &  7.990$+$06  &   2.696$-$04  &   3.035$+$06  &   9.957$-$05  &   2.71  \\
    1  &   12   &  2.844$+$05  &   4.787$-$06  &   6.461$+$05  &   1.065$-$05  &   0.45  \\
    1  &   13   &  2.573$+$08  &   4.192$-$03  &   3.068$+$08  &   4.896$-$03  &   0.86  \\
\hline  											       								   					      
\end{longtable}

                               
\begin{flushleft}                                                                               
                               
{\small
                                                                               
}                                                                                               
                               
\end{flushleft} 


\TableExplanation

\bigskip
\renewcommand{\arraystretch}{1.0}

\section*{Table 1.\label{tbl1te} Energies (Ryd) for the lowest 375 levels of Sr~IV and their lifetimes ($\tau$, s). }
\begin{tabular}{@{}p{1in}p{6in}@{}}
Index            & Level Index \\
Configuration    & The configuration to which the level belongs \\
Level             & The $LSJ$ designation of the level \\
Energy           & Present energies from the {\sc grasp} code  with 3990 level calculations \\
$\tau$ (s)       & Lifetime of the level in s \\

\end{tabular}
\label{tableII}

\bigskip
\renewcommand{\arraystretch}{1.0}

\section*{Table 2.\label{tbl2te} Energies (Ryd) for the lowest 375 levels of Y~V and their lifetimes ($\tau$, s). }
\begin{tabular}{@{}p{1in}p{6in}@{}}
Index            & Level Index \\
Configuration    & The configuration to which the level belongs \\
Level             & The $LSJ$ designation of the level \\
Energy           & Present energies from the {\sc grasp} code  with 3990 level calculations \\
$\tau$ (s)       & Lifetime of the level in s \\

\end{tabular}
\label{tableII}

\bigskip
\renewcommand{\arraystretch}{1.0}

\section*{Table 3.\label{tbl3te} Energies (Ryd) for the lowest 375 levels of Zr~VI and their lifetimes ($\tau$, s). }
\begin{tabular}{@{}p{1in}p{6in}@{}}
Index            & Level Index \\
Configuration    & The configuration to which the level belongs \\
Level             & The $LSJ$ designation of the level \\
Energy           & Present energies from the {\sc grasp} code  with 3990 level calculations \\
$\tau$ (s)       & Lifetime of the level in s \\

\end{tabular}
\label{tableII}

\bigskip
\renewcommand{\arraystretch}{1.0}

\section*{Table 4.\label{tbl4te} Energies (Ryd) for the lowest 375 levels of Nb~VII and their lifetimes ($\tau$, s). }
\begin{tabular}{@{}p{1in}p{6in}@{}}
Index            & Level Index \\
Configuration    & The configuration to which the level belongs \\
Level             & The $LSJ$ designation of the level \\
Energy           & Present energies from the {\sc grasp} code  with 3990 level calculations \\
$\tau$ (s)       & Lifetime of the level in s \\

\end{tabular}
\label{tableII}

\bigskip
\renewcommand{\arraystretch}{1.0}

\section*{Table 5.\label{tbl5te} Energies (Ryd) for the lowest 375 levels of Mo~VIII and their lifetimes ($\tau$, s). }
\begin{tabular}{@{}p{1in}p{6in}@{}}
Index            & Level Index \\
Configuration    & The configuration to which the level belongs \\
Level             & The $LSJ$ designation of the level \\
Energy           & Present energies from the {\sc grasp} code  with 3990 level calculations \\
$\tau$ (s)       & Lifetime of the level in s \\

\end{tabular}
\label{tableII}


\bigskip
\section*{Table 6.\label{tbl6te}  Transition wavelengths ($\lambda_{ij}$ in $\rm \AA$), radiative rates (A$_{ji}$ in s$^{-1}$),
 oscillator strengths (f$_{ij}$, dimensionless), and line strengths (S, in atomic units) for electric dipole (E1), and 
A$_{ji}$ for electric quadrupole (E2), magnetic dipole (M1), and magnetic quadrupole (M2) transitions of Sr~IV.  The ratio R(E1) of 
velocity and length forms of A- values for E1 transitions is listed in the last column.}
\begin{tabular}{@{}p{1in}p{6in}@{}}
$i$ and $j$         & The lower ($i$) and upper ($j$) levels of a transition as defined in Table 1.\\
$\lambda_{ij}$      & Transition wavelength (in ${\rm \AA}$) \\
A$^{E1}_{ji}$       & Radiative transition probability (in s$^{-1}$) for the E1 transitions \\
f$^{E1}_{ij}$       & Absorption oscillator strength (dimensionless) for the E1 transitions \\
S$^{E1}$            & Line strength in atomic unit (a.u.), 1 a.u. = 6.460$\times$10$^{-36}$ cm$^2$ esu$^2$ for the E1 transitions \\
A$^{E2}_{ji}$       & Radiative transition probability (in s$^{-1}$) for the E2 transitions \\
A$^{M1}_{ji}$       & Radiative transition probability (in s$^{-1}$) for the M1 transitions \\
A$^{M2}_{ji}$       & Radiative transition probability (in s$^{-1}$) for the M2 transitions \\
R(E1)                     & Ratio of velocity and length forms of A- (or f- and S-) values for the E1 transitions \\
$a{\pm}b$ &  $\equiv a\times{10^{{\pm}b}}$ \\
\end{tabular}
\label{ExplTable6}

\bigskip
\section*{Table 7.\label{tbl7te}  Transition wavelengths ($\lambda_{ij}$ in $\rm \AA$), radiative rates (A$_{ji}$ in s$^{-1}$),
 oscillator strengths (f$_{ij}$, dimensionless), and line strengths (S, in atomic units) for electric dipole (E1), and 
A$_{ji}$ for electric quadrupole (E2), magnetic dipole (M1), and magnetic quadrupole (M2) transitions of Y~V.   The ratio R(E1) of 
velocity and length forms of A- values for E1 transitions is listed in the last column.}
\begin{tabular}{@{}p{1in}p{6in}@{}}
$i$ and $j$         & The lower ($i$) and upper ($j$) levels of a transition as defined in Table 2.\\
$\lambda_{ij}$      & Transition wavelength (in ${\rm \AA}$) \\
A$^{E1}_{ji}$       & Radiative transition probability (in s$^{-1}$) for the E1 transitions \\
f$^{E1}_{ij}$       & Absorption oscillator strength (dimensionless) for the E1 transitions \\
S$^{E1}$            & Line strength in atomic unit (a.u.), 1 a.u. = 6.460$\times$10$^{-36}$ cm$^2$ esu$^2$ for the E1 transitions \\
A$^{E2}_{ji}$       & Radiative transition probability (in s$^{-1}$) for the E2 transitions \\
A$^{M1}_{ji}$       & Radiative transition probability (in s$^{-1}$) for the M1 transitions \\
A$^{M2}_{ji}$       & Radiative transition probability (in s$^{-1}$) for the M2 transitions \\
R(E1)                     & Ratio of velocity and length forms of A- (or f- and S-) values for the E1 transitions \\
$a{\pm}b$ &  $\equiv a\times{10^{{\pm}b}}$ \\
\end{tabular}
\label{ExplTable7}

\bigskip
\section*{Table 8.\label{tbl8te}  Transition wavelengths ($\lambda_{ij}$ in $\rm \AA$), radiative rates (A$_{ji}$ in s$^{-1}$),
 oscillator strengths (f$_{ij}$, dimensionless), and line strengths (S, in atomic units) for electric dipole (E1), and 
A$_{ji}$ for electric quadrupole (E2), magnetic dipole (M1), and magnetic quadrupole (M2) transitions of Zr~VI.   The ratio R(E1) of 
velocity and length forms of A- values for E1 transitions is listed in the last column.}
\begin{tabular}{@{}p{1in}p{6in}@{}}
$i$ and $j$         & The lower ($i$) and upper ($j$) levels of a transition as defined in Table 3.\\
$\lambda_{ij}$      & Transition wavelength (in ${\rm \AA}$) \\
A$^{E1}_{ji}$       & Radiative transition probability (in s$^{-1}$) for the E1 transitions \\
f$^{E1}_{ij}$       & Absorption oscillator strength (dimensionless) for the E1 transitions \\
S$^{E1}$            & Line strength in atomic unit (a.u.), 1 a.u. = 6.460$\times$10$^{-36}$ cm$^2$ esu$^2$ for the E1 transitions \\
A$^{E2}_{ji}$       & Radiative transition probability (in s$^{-1}$) for the E2 transitions \\
A$^{M1}_{ji}$       & Radiative transition probability (in s$^{-1}$) for the M1 transitions \\
A$^{M2}_{ji}$       & Radiative transition probability (in s$^{-1}$) for the M2 transitions \\
R(E1)                     & Ratio of velocity and length forms of A- (or f- and S-) values for the E1 transitions \\
$a{\pm}b$ &  $\equiv a\times{10^{{\pm}b}}$ \\
\end{tabular}
\label{ExplTable8}

\bigskip
\section*{Table 9.\label{tbl9te}  Transition wavelengths ($\lambda_{ij}$ in $\rm \AA$), radiative rates (A$_{ji}$ in s$^{-1}$),
 oscillator strengths (f$_{ij}$, dimensionless), and line strengths (S, in atomic units) for electric dipole (E1), and 
A$_{ji}$ for electric quadrupole (E2), magnetic dipole (M1), and magnetic quadrupole (M2) transitions of Nb~VII.   The ratio R(E1) of 
velocity and length forms of A- values for E1 transitions is listed in the last column.}
\begin{tabular}{@{}p{1in}p{6in}@{}}
$i$ and $j$         & The lower ($i$) and upper ($j$) levels of a transition as defined in Table 4.\\
$\lambda_{ij}$      & Transition wavelength (in ${\rm \AA}$) \\
A$^{E1}_{ji}$       & Radiative transition probability (in s$^{-1}$) for the E1 transitions \\
f$^{E1}_{ij}$       & Absorption oscillator strength (dimensionless) for the E1 transitions \\
S$^{E1}$            & Line strength in atomic unit (a.u.), 1 a.u. = 6.460$\times$10$^{-36}$ cm$^2$ esu$^2$ for the E1 transitions \\
A$^{E2}_{ji}$       & Radiative transition probability (in s$^{-1}$) for the E2 transitions \\
A$^{M1}_{ji}$       & Radiative transition probability (in s$^{-1}$) for the M1 transitions \\
A$^{M2}_{ji}$       & Radiative transition probability (in s$^{-1}$) for the M2 transitions \\
R(E1)                     & Ratio of velocity and length forms of A- (or f- and S-) values for the E1 transitions \\
$a{\pm}b$ &  $\equiv a\times{10^{{\pm}b}}$ \\
\end{tabular}
\label{ExplTable9}

\bigskip
\section*{Table 10.\label{tbl10te}  Transition wavelengths ($\lambda_{ij}$ in $\rm \AA$), radiative rates (A$_{ji}$ in s$^{-1}$),
 oscillator strengths (f$_{ij}$, dimensionless), and line strengths (S, in atomic units) for electric dipole (E1), and 
A$_{ji}$ for electric quadrupole (E2), magnetic dipole (M1), and magnetic quadrupole (M2) transitions of Mo~VIII.   The ratio R(E1) of 
velocity and length forms of A- values for E1 transitions is listed in the last column.}
\begin{tabular}{@{}p{1in}p{6in}@{}}
$i$ and $j$         & The lower ($i$) and upper ($j$) levels of a transition as defined in Table 5.\\
$\lambda_{ij}$      & Transition wavelength (in ${\rm \AA}$) \\
A$^{E1}_{ji}$       & Radiative transition probability (in s$^{-1}$) for the E1 transitions \\
f$^{E1}_{ij}$       & Absorption oscillator strength (dimensionless) for the E1 transitions \\
S$^{E1}$            & Line strength in atomic unit (a.u.), 1 a.u. = 6.460$\times$10$^{-36}$ cm$^2$ esu$^2$ for the E1 transitions \\
A$^{E2}_{ji}$       & Radiative transition probability (in s$^{-1}$) for the E2 transitions \\
A$^{M1}_{ji}$       & Radiative transition probability (in s$^{-1}$) for the M1 transitions \\
A$^{M2}_{ji}$       & Radiative transition probability (in s$^{-1}$) for the M2 transitions \\
R(E1)                     & Ratio of velocity and length forms of A- (or f- and S-) values for the E1 transitions \\
$a{\pm}b$ &  $\equiv a\times{10^{{\pm}b}}$ \\
\end{tabular}
\label{ExplTable10}

\datatables 



\setlength{\LTleft}{0pt}
\setlength{\LTright}{0pt} 


\setlength{\tabcolsep}{0.5\tabcolsep}

\renewcommand{\arraystretch}{1.0}

\footnotesize 

\begin{longtable}{@{\extracolsep\fill}rllll@{}}
\caption{Energies (Ryd) for the lowest 375 levels of Sr IV and their lifetimes ($\tau$, s). $a{\pm}b \equiv a{\times}$10$^{{\pm}b}$. See page\ \pageref{tbl1te} for Explanation of Tables}
Index  & Configuration                       & Level              & Energy  & $\tau$  (s)   \\
\hline\\
\endfirsthead\\
\caption[]{(continued)}
Index  & Configuration                       & Level              & Energy  & $\tau$  (s)   \\
\hline\\
\endhead
    1 &   4s$^2$4p$^5$  		   &  $^2$P$^o_{3/2 }$    &  0.00000  &   .......... \\	
    2 &   4s$^2$4p$^5$  		   &  $^2$P$^o_{1/2 }$    &  0.08515  &   6.816$-$02 \\      
    3 &   4s4p$^6$			   &  $^2$S$  _{1/2 }$    &  1.44244  &   1.897$-$09 \\  
    4 &   4s$^2$4p$^4$($^3$P)4d 	   &  $^4$D$  _{7/2 }$    &  1.75452  &   1.196$-$01 \\  
    5 &   4s$^2$4p$^4$($^3$P)4d 	   &  $^4$D$  _{5/2 }$    &  1.75612  &   9.770$-$08 \\  
    6 &   4s$^2$4p$^4$($^3$P)4d 	   &  $^4$D$  _{3/2 }$    &  1.76432  &   3.506$-$07 \\  
    7 &   4s$^2$4p$^4$($^3$P)4d 	   &  $^4$D$  _{1/2 }$    &  1.77444  &   1.272$-$06 \\  
    8 &   4s$^2$4p$^4$($^3$P)4d 	   &  $^4$F$  _{9/2 }$    &  1.85784  &   7.869$-$01 \\
\end{longtable}

\clearpage
\newpage

\begin{longtable}{@{\extracolsep\fill}rllll@{}}
\caption{Energies (Ryd) for the lowest 375 levels of Y V and their lifetimes ($\tau$, s). $a{\pm}b \equiv a{\times}$10$^{{\pm}b}$. See page\ \pageref{tbl2te} for Explanation of Tables}
Index  & Configuration                       & Level              & Energy  & $\tau$  (s)   \\
\hline\\
\endfirsthead\\
\caption[]{(continued)}
Index  & Configuration                       & Level              & Energy  & $\tau$  (s)   \\
\hline\\
\endhead
    1  & 4s$^2$4p$^5$			    &  $^2$P$^o_{3/2 }$        &     0.00000  &   .......... \\
    2  & 4s$^2$4p$^5$			    &  $^2$P$^o_{1/2 }$        &     0.10989  &   3.171$-$02 \\
    3  & 4s4p$^6$			    &  $^2$S$  _{1/2 }$        &     1.64719  &   1.011$-$09 \\
    4  & 4s$^2$4p$^4$($^3$P)4d  	    &  $^4$D$  _{5/2 }$        &     2.04945  &   6.765$-$08 \\
    5  & 4s$^2$4p$^4$($^3$P)4d  	    &  $^4$D$  _{7/2 }$        &     2.04923  &   6.441$-$02 \\
    6  & 4s$^2$4p$^4$($^3$P)4d  	    &  $^4$D$  _{3/2 }$        &     2.05923  &   1.654$-$07 \\
    7  & 4s$^2$4p$^4$($^3$P)4d  	    &  $^4$D$  _{1/2 }$        &     2.07282  &   2.471$-$07 \\
    8  & 4s$^2$4p$^4$($^3$P)4d  	    &  $^4$F$  _{9/2 }$        &     2.16295  &   4.710$-$01 \\
\end{longtable}

\clearpage
\newpage

\begin{longtable}{@{\extracolsep\fill}rllll@{}}
\caption{Energies (Ryd) for the lowest 375 levels of Zr VI and their lifetimes ($\tau$, s). $a{\pm}b \equiv a{\times}$10$^{{\pm}b}$. See page\ \pageref{tbl3te} for Explanation of Tables}
Index  & Configuration                       & Level              & Energy  & $\tau$  (s)   \\
\hline\\
\endfirsthead\\
\caption[]{(continued)}
Index  & Configuration                       & Level              & Energy  & $\tau$  (s)   \\
\hline\\
\endhead
    1    &    4s$^2$4p$^5$			 &  $^2$P$^o_{3/2 }$	 & 0.00000  &	..........    \\
    2    &    4s$^2$4p$^5$			 &  $^2$P$^o_{1/2 }$	 & 0.13831  &	1.591$-$02    \\
    3    &    4s4p$^6$  			 &  $^2$S$  _{1/2 }$	 & 1.85108  &	6.388$-$10    \\
    4    &    4s$^2$4p$^4$($^3$P)4d		 &  $^4$D$  _{5/2 }$	 & 2.33330  &	4.910$-$08    \\
    5    &    4s$^2$4p$^4$($^3$P)4d		 &  $^4$D$  _{7/2 }$	 & 2.33503  &	3.883$-$02    \\
    6    &    4s$^2$4p$^4$($^3$P)4d		 &  $^4$D$  _{3/2 }$	 & 2.34467  &	8.945$-$08    \\
    7    &    4s$^2$4p$^4$($^3$P)4d		 &  $^4$D$  _{1/2 }$	 & 2.36246  &	9.255$-$08    \\
    8    &    4s$^2$4p$^4$($^3$P)4d		 &  $^4$F$  _{9/2 }$	 & 2.45699  &	3.080$-$01    \\
\end{longtable}

\clearpage
\newpage

\begin{longtable}{@{\extracolsep\fill}rllll@{}}
\caption{Energies (Ryd) for the lowest 375 levels of Nb VII and their lifetimes ($\tau$, s). $a{\pm}b \equiv a{\times}$10$^{{\pm}b}$. See page\ \pageref{tbl4te} for Explanation of Tables}
Index  & Configuration                       & Level              & Energy  & $\tau$  (s)   \\
\hline\\
\endfirsthead\\
\caption[]{(continued)}
Index  & Configuration                       & Level              & Energy  & $\tau$  (s)   \\
\hline\\
\endhead
    1   &   4s$^2$4p$^5$		      &   $^2$P$^o_{3/2 }$     & 0.00000   &   ..........    \\
    2   &   4s$^2$4p$^5$		      &   $^2$P$^o_{1/2 }$     & 0.17076   &   8.452$-$03    \\
    3   &   4s4p$^6$			      &   $^2$S$  _{1/2 }$     & 2.05605   &   4.422$-$10    \\
    4   &   4s$^2$4p$^4$($^3$P)4d	      &   $^4$D$  _{5/2 }$     & 2.61194   &   3.708$-$08    \\
    5   &   4s$^2$4p$^4$($^3$P)4d	      &   $^4$D$  _{7/2 }$     & 2.61624   &   2.526$-$02    \\
    6   &   4s$^2$4p$^4$($^3$P)4d	      &   $^4$D$  _{3/2 }$     & 2.62476   &   5.270$-$08    \\
    7   &   4s$^2$4p$^4$($^3$P)4d	      &   $^4$D$  _{1/2 }$     & 2.64747   &   4.359$-$08    \\
    8   &   4s$^2$4p$^4$($^3$P)4d	      &   $^4$F$  _{9/2 }$     & 2.74502   &   2.149$-$01    \\
\end{longtable}

\clearpage
\newpage

\begin{longtable}{@{\extracolsep\fill}rllll@{}}
\caption{Energies (Ryd) for the lowest 375 levels of Mo VIII and their lifetimes ($\tau$, s). $a{\pm}b \equiv a{\times}$10$^{{\pm}b}$. See page\ \pageref{tbl5te} for Explanation of Tables}
Index  & Configuration                       & Level              & Energy  & $\tau$  (s)   \\
\hline\\
\endfirsthead\\
\caption[]{(continued)}
Index  & Configuration                       & Level              & Energy  & $\tau$  (s)   \\
\hline\\
\endhead
    1  &  4s$^2$4p$^5$  		     &   $^2$P$^o_{3/2 }$     &  0.00000    &	..........    \\
    2  &  4s$^2$4p$^5$  		     &   $^2$P$^o_{1/2 }$     &  0.20760    &	4.703$-$03    \\
    3  &  4s4p$^6$			     &   $^2$S$  _{1/2 }$     &  2.26319    &	3.251$-$10    \\
    4  &  4s$^2$4p$^4$($^3$P)4d 	     &   $^4$D$  _{5/2 }$     &  2.88728    &	2.896$-$08    \\
    5  &  4s$^2$4p$^4$($^3$P)4d 	     &   $^4$D$  _{7/2 }$     &  2.89480    &	1.737$-$02    \\
    6  &  4s$^2$4p$^4$($^3$P)4d 	     &   $^4$D$  _{3/2 }$     &  2.90130    &	3.320$-$08    \\
    7  &  4s$^2$4p$^4$($^3$P)4d 	     &   $^4$D$  _{1/2 }$     &  2.92955    &	2.339$-$08    \\
    8  &  4s$^2$4p$^4$($^3$P)4d 	     &   $^4$F$  _{9/2 }$     &  3.02938    &	1.576$-$01    \\
\end{longtable}

\newpage

\renewcommand{\arraystretch}{1.0}

\footnotesize 

\begin{longtable}{@{\extracolsep\fill}rrrrrrrrrr@{}}
\caption{Transition wavelengths ($\lambda_{ij}$ in $\rm \AA$), radiative rates (A$_{ji}$ in s$^{-1}$), oscillator strengths (f$_{ij}$, dimensionless), and line     
strengths (S, in atomic units) for electric dipole (E1), and A$_{ji}$ for E2, M1, and M2 transitions in Sr~IV.  The last column gives R(E1), the ratio of velocity and length forms of A- values for E1 transitions. $a{\pm}b \equiv a{\times}$10$^{{\pm}b}$. See page\ \pageref{tbl6te} for Explanation of Tables}
$i$ & $j$ & $\lambda_{ij}$ & A$^{{\rm E1}}_{ji}$  & f$^{{\rm E1}}_{ij}$ & S$^{{\rm E1}}$ & A$^{{\rm E2}}_{ji}$  & A$^{{\rm M1}}_{ji}$ & A$^{{\rm M2}}_{ji}$ & R(E1)  \\
\hline\\
\endfirsthead\\
\caption[]{(continued)}
$i$ & $j$ & $\lambda_{ij}$ & A$^{{\rm E1}}_{ji}$  & f$^{{\rm E1}}_{ij}$ & S$^{{\rm E1}}$ & A$^{{\rm E2}}_{ji}$  & A$^{{\rm M1}}_{ji}$ & A$^{{\rm M2}}_{ji}$  & R(E1) \\
\hline\\
\endhead
    1 &    2 &  1.070$+$04 &  0.000$+$00 &  0.000$+$00 &  0.000$+$00 &  1.735$-$02 &  1.465$+$01 &  0.000$+$00 &   0.0$+$00 \\
    1 &    3 &  6.318$+$02 &  3.546$+$08 &  1.061$-$02 &  8.826$-$02 &  0.000$+$00 &  0.000$+$00 &  6.059$+$00 &   2.3$-$01 \\
    1 &    4 &  5.194$+$02 &  0.000$+$00 &  0.000$+$00 &  0.000$+$00 &  0.000$+$00 &  0.000$+$00 &  8.362$+$00 &   0.0$+$00 \\
    1 &    5 &  5.189$+$02 &  1.024$+$07 &  6.198$-$04 &  4.235$-$03 &  0.000$+$00 &  0.000$+$00 &  1.837$-$01 &   9.5$-$01 \\
    1 &    6 &  5.165$+$02 &  2.556$+$06 &  1.022$-$04 &  6.954$-$04 &  0.000$+$00 &  0.000$+$00 &  1.370$-$01 &   7.8$-$01 \\
    1 &    7 &  5.135$+$02 &  4.257$+$05 &  8.415$-$06 &  5.691$-$05 &  0.000$+$00 &  0.000$+$00 &  1.152$+$00 &   4.8$-$01 \\
    1 &    9 &  4.821$+$02 &  0.000$+$00 &  0.000$+$00 &  0.000$+$00 &  0.000$+$00 &  0.000$+$00 &  7.617$-$01 &   0.0$+$00 \\
    1 &   10 &  4.765$+$02 &  8.106$+$07 &  4.138$-$03 &  2.597$-$02 &  0.000$+$00 &  0.000$+$00 &  7.902$-$02 &   9.2$-$01 \\
\end{longtable}


\renewcommand{\arraystretch}{1.0}

\footnotesize 

\begin{longtable}{@{\extracolsep\fill}rrrrrrrrrr@{}}
\caption{Transition wavelengths ($\lambda_{ij}$ in $\rm \AA$), radiative rates (A$_{ji}$ in s$^{-1}$), oscillator strengths (f$_{ij}$, dimensionless), and line     
strengths (S, in atomic units) for electric dipole (E1), and A$_{ji}$ for E2, M1, and M2 transitions in Y~V. The last column gives R(E1), the ratio of velocity and length forms of A- values for E1 transitions.   $a{\pm}b \equiv a{\times}$10$^{{\pm}b}$. See page\ \pageref{tbl7te} for Explanation of Tables}
$i$ & $j$ & $\lambda_{ij}$ & A$^{{\rm E1}}_{ji}$  & f$^{{\rm E1}}_{ij}$ & S$^{{\rm E1}}$ & A$^{{\rm E2}}_{ji}$  & A$^{{\rm M1}}_{ji}$ & A$^{{\rm M2}}_{ji}$ & R(E1)  \\
\hline\\
\endfirsthead\\
\caption[]{(continued)}
$i$ & $j$ & $\lambda_{ij}$ & A$^{{\rm E1}}_{ji}$  & f$^{{\rm E1}}_{ij}$ & S$^{{\rm E1}}$ & A$^{{\rm E2}}_{ji}$  & A$^{{\rm M1}}_{ji}$ & A$^{{\rm M2}}_{ji}$ & R(E1)  \\
\hline\\
\endhead
    1 &    2 &  8.292$+$03 &  0.000$+$00 &  0.000$+$00 &  0.000$+$00 &  4.269$-$02 &  3.149$+$01 &  0.000$+$00 &   0.0$+$00 \\
    1 &    3 &  5.532$+$02 &  6.761$+$08 &  1.551$-$02 &  1.130$-$01 &  0.000$+$00 &  0.000$+$00 &  1.012$+$01 &   3.6$-$01 \\
    1 &    4 &  4.446$+$02 &  1.478$+$07 &  6.572$-$04 &  3.848$-$03 &  0.000$+$00 &  0.000$+$00 &  3.417$-$01 &   9.3$-$01 \\
    1 &    5 &  4.447$+$02 &  0.000$+$00 &  0.000$+$00 &  0.000$+$00 &  0.000$+$00 &  0.000$+$00 &  1.553$+$01 &   0.0$+$00 \\
    1 &    6 &  4.425$+$02 &  5.943$+$06 &  1.745$-$04 &  1.017$-$03 &  0.000$+$00 &  0.000$+$00 &  2.254$-$01 &   9.0$-$01 \\
    1 &    7 &  4.396$+$02 &  2.088$+$06 &  3.025$-$05 &  1.751$-$04 &  0.000$+$00 &  0.000$+$00 &  2.553$+$00 &   8.4$-$01 \\
    1 &    9 &  4.137$+$02 &  0.000$+$00 &  0.000$+$00 &  0.000$+$00 &  0.000$+$00 &  0.000$+$00 &  1.855$+$00 &   0.0$+$00 \\
    1 &   10 &  4.083$+$02 &  1.311$+$08 &  4.915$-$03 &  2.643$-$02 &  0.000$+$00 &  0.000$+$00 &  2.078$-$01 &   9.3$-$01 \\
\end{longtable}


\renewcommand{\arraystretch}{1.0}

\footnotesize 

\begin{longtable}{@{\extracolsep\fill}rrrrrrrrrr@{}}
\caption{Transition wavelengths ($\lambda_{ij}$ in $\rm \AA$), radiative rates (A$_{ji}$ in s$^{-1}$), oscillator strengths (f$_{ij}$, dimensionless), and line     
strengths (S, in atomic units) for electric dipole (E1), and A$_{ji}$ for E2, M1, and M2 transitions in Zr~VI.  The last column gives R(E1), the ratio of velocity and length forms of A- values for E1 transitions.  $a{\pm}b \equiv a{\times}$10$^{{\pm}b}$. See page\ \pageref{tbl8te} for Explanation of Tables}
$i$ & $j$ & $\lambda_{ij}$ & A$^{{\rm E1}}_{ji}$  & f$^{{\rm E1}}_{ij}$ & S$^{{\rm E1}}$ & A$^{{\rm E2}}_{ji}$  & A$^{{\rm M1}}_{ji}$ & A$^{{\rm M2}}_{ji}$ & R(E1)  \\
\hline\\
\endfirsthead\\
\caption[]{(continued)}
$i$ & $j$ & $\lambda_{ij}$ & A$^{{\rm E1}}_{ji}$  & f$^{{\rm E1}}_{ij}$ & S$^{{\rm E1}}$ & A$^{{\rm E2}}_{ji}$  & A$^{{\rm M1}}_{ji}$ & A$^{{\rm M2}}_{ji}$  & R(E1) \\
\hline\\
\endhead
    1 &    2 &  6.589$+$03 &  0.000$+$00 &  0.000$+$00 &  0.000$+$00 &  9.836$-$02 &  6.277$+$01 &  0.000$+$00 &   0.0$+$00 \\
    1 &    3 &  4.923$+$02 &  1.081$+$09 &  1.964$-$02 &  1.273$-$01 &  0.000$+$00 &  0.000$+$00 &  1.579$+$01 &   4.2$-$01 \\
    1 &    4 &  3.905$+$02 &  2.037$+$07 &  6.986$-$04 &  3.593$-$03 &  0.000$+$00 &  0.000$+$00 &  5.571$-$01 &   9.2$-$01 \\
    1 &    5 &  3.903$+$02 &  0.000$+$00 &  0.000$+$00 &  0.000$+$00 &  0.000$+$00 &  0.000$+$00 &  2.575$+$01 &   0.0$+$00 \\
    1 &    6 &  3.887$+$02 &  1.118$+$07 &  2.532$-$04 &  1.296$-$03 &  0.000$+$00 &  0.000$+$00 &  3.260$-$01 &   9.6$-$01 \\
    1 &    7 &  3.857$+$02 &  5.771$+$06 &  6.436$-$05 &  3.269$-$04 &  0.000$+$00 &  0.000$+$00 &  5.114$+$00 &   9.6$-$01 \\
    1 &    9 &  3.639$+$02 &  0.000$+$00 &  0.000$+$00 &  0.000$+$00 &  0.000$+$00 &  0.000$+$00 &  3.938$+$00 &   0.0$+$00 \\
    1 &   10 &  3.592$+$02 &  1.219$+$07 &  1.179$-$04 &  5.579$-$04 &  0.000$+$00 &  0.000$+$00 &  3.468$+$01 &   1.4$+$00 \\
\end{longtable}

\newpage

\renewcommand{\arraystretch}{1.0}

\footnotesize 

\begin{longtable}{@{\extracolsep\fill}rrrrrrrrrr@{}}
\caption{Transition wavelengths ($\lambda_{ij}$ in $\rm \AA$), radiative rates (A$_{ji}$ in s$^{-1}$), oscillator strengths (f$_{ij}$, dimensionless), and line     
strengths (S, in atomic units) for electric dipole (E1), and A$_{ji}$ for E2, M1, and M2 transitions in Nb~VII.  The last column gives R(E1), the ratio of velocity and length forms of A- values for E1 transitions.  $a{\pm}b \equiv a{\times}$10$^{{\pm}b}$. See page\ \pageref{tbl9te} for Explanation of Tables}
$i$ & $j$ & $\lambda_{ij}$ & A$^{{\rm E1}}_{ji}$  & f$^{{\rm E1}}_{ij}$ & S$^{{\rm E1}}$ & A$^{{\rm E2}}_{ji}$  & A$^{{\rm M1}}_{ji}$ & A$^{{\rm M2}}_{ji}$ & R(E1)  \\
\hline\\
\endfirsthead\\
\caption[]{(continued)}
$i$ & $j$ & $\lambda_{ij}$ & A$^{{\rm E1}}_{ji}$  & f$^{{\rm E1}}_{ij}$ & S$^{{\rm E1}}$ & A$^{{\rm E2}}_{ji}$  & A$^{{\rm M1}}_{ji}$ & A$^{{\rm M2}}_{ji}$  & R(E1) \\
\hline\\
\endhead
    1 &    2 &  5.337$+$03 &  0.000$+$00 &  0.000$+$00 &  0.000$+$00 &  2.150$-$01 &  1.181$+$02 &  0.000$+$00 &   0.0$+$00 \\
    1 &    3 &  4.432$+$02 &  1.574$+$09 &  2.318$-$02 &  1.353$-$01 &  0.000$+$00 &  0.000$+$00 &  2.345$+$01 &   4.4$-$01 \\
    1 &    4 &  3.489$+$02 &  2.697$+$07 &  7.383$-$04 &  3.392$-$03 &  0.000$+$00 &  0.000$+$00 &  8.343$-$01 &   9.1$-$01 \\
    1 &    5 &  3.483$+$02 &  0.000$+$00 &  0.000$+$00 &  0.000$+$00 &  0.000$+$00 &  0.000$+$00 &  3.959$+$01 &   0.0$+$00 \\
    1 &    6 &  3.472$+$02 &  1.886$+$07 &  3.409$-$04 &  1.558$-$03 &  0.000$+$00 &  0.000$+$00 &  4.284$-$01 &   9.8$-$01 \\
    1 &    7 &  3.442$+$02 &  1.293$+$07 &  1.148$-$04 &  5.203$-$04 &  0.000$+$00 &  0.000$+$00 &  9.512$+$00 &   9.8$-$01 \\
    1 &    9 &  3.255$+$02 &  0.000$+$00 &  0.000$+$00 &  0.000$+$00 &  0.000$+$00 &  0.000$+$00 &  7.463$+$00 &   0.0$+$00 \\
    1 &   10 &  3.216$+$02 &  2.169$+$07 &  1.681$-$04 &  7.120$-$04 &  0.000$+$00 &  0.000$+$00 &  4.979$+$01 &   1.4$+$00 \\
\end{longtable}


\renewcommand{\arraystretch}{1.0}

\footnotesize 

\begin{longtable}{@{\extracolsep\fill}rrrrrrrrrr@{}}
\caption{Transition wavelengths ($\lambda_{ij}$ in $\rm \AA$), radiative rates (A$_{ji}$ in s$^{-1}$), oscillator strengths (f$_{ij}$, dimensionless), and line     
strengths (S, in atomic units) for electric dipole (E1), and A$_{ji}$ for E2, M1, and M2 transitions in Mo~VIII.  The last column gives R(E1), the ratio of velocity and length forms of A- values for E1 transitions.  $a{\pm}b \equiv a{\times}$10$^{{\pm}b}$. See page\ \pageref{tbl10te} for Explanation of Tables}
$i$ & $j$ & $\lambda_{ij}$ & A$^{{\rm E1}}_{ji}$  & f$^{{\rm E1}}_{ij}$ & S$^{{\rm E1}}$ & A$^{{\rm E2}}_{ji}$  & A$^{{\rm M1}}_{ji}$ & A$^{{\rm M2}}_{ji}$ & R(E1)  \\
\hline\\
\endfirsthead\\
\caption[]{(continued)}
$i$ & $j$ & $\lambda_{ij}$ & A$^{{\rm E1}}_{ji}$  & f$^{{\rm E1}}_{ij}$ & S$^{{\rm E1}}$ & A$^{{\rm E2}}_{ji}$  & A$^{{\rm M1}}_{ji}$ & A$^{{\rm M2}}_{ji}$ & R(E1)  \\
\hline\\
\endhead
    1 &    2 &  4.389$+$03 &  0.000$+$00 &  0.000$+$00 &  0.000$+$00 &  4.485$-$01 &  2.122$+$02 &  0.000$+$00 &   0.0$+$00 \\
    1 &    3 &  4.026$+$02 &  2.157$+$09 &  2.621$-$02 &  1.390$-$01 &  0.000$+$00 &  0.000$+$00 &  3.360$+$01 &   4.5$-$01 \\
    1 &    4 &  3.156$+$02 &  3.453$+$07 &  7.734$-$04 &  3.214$-$03 &  0.000$+$00 &  0.000$+$00 &  1.176$+$00 &   9.0$-$01 \\
    1 &    5 &  3.148$+$02 &  0.000$+$00 &  0.000$+$00 &  0.000$+$00 &  0.000$+$00 &  0.000$+$00 &  5.756$+$01 &   0.0$+$00 \\
    1 &    6 &  3.141$+$02 &  2.953$+$07 &  4.367$-$04 &  1.806$-$03 &  0.000$+$00 &  0.000$+$00 &  5.210$-$01 &   9.8$-$01 \\
    1 &    7 &  3.111$+$02 &  2.562$+$07 &  1.859$-$04 &  7.613$-$04 &  0.000$+$00 &  0.000$+$00 &  1.662$+$01 &   9.7$-$01 \\
    1 &    9 &  2.949$+$02 &  0.000$+$00 &  0.000$+$00 &  0.000$+$00 &  0.000$+$00 &  0.000$+$00 &  1.285$+$01 &   0.0$+$00 \\
    1 &   10 &  2.912$+$02 &  2.995$+$07 &  1.903$-$04 &  7.298$-$04 &  0.000$+$00 &  0.000$+$00 &  6.731$+$01 &   1.3$+$00 \\
\end{longtable}

\end{document}